\documentclass[12pt,preprint]{aastex631}

\usepackage{amsmath}
\usepackage{graphicx}

\newcommand{\bu}{{\bf u}}

\newcommand{\bnabla}{{\bf \nabla}}
\renewcommand{\Pr}{\mathrm{Pr}}

\newcommand\dogout{\bgroup\markoverwith{\textcolor{red}{\rule[0.5ex]{2pt}{1pt}}}\ULon}
\newcommand\pgout{\bgroup\markoverwith{\textcolor{blue}{\rule[0.5ex]{2pt}{1pt}}}\ULon}
\newcommand\lmout{\bgroup\markoverwith{\textcolor{orange}{\rule[0.5ex]{2pt}{1pt}}}\ULon}

\begin{document}

\title{Towards a self-consistent hydrodynamical model of the solar tachocline}
\shorttitle{The solar tachocline}

\author{P. Garaud}
\affil{
Department of Applied Mathematics, Baskin School of Engineering, \\
University of California, Santa Cruz, CA 95064, USA
}

\author{D. O. Gough}
\affil{
Institute of Astronomy, University of Cambridge, Madingley Road, 
Cambridge
CB3 0HA} 
\affil{Department of Applied Mathematics and Theoretical Physics, Centre for Mathematical Sciences, \\ University of Cambridge, Wilberforce Road, Cambridge CB3 0WA }

\author{L. I. Matilsky$^1$}

\shortauthors{Garaud, Gough \& Matilsky}


\begin{abstract}
The solar tachocline is an internal boundary layer in the Sun located between the differentially-rotating convection zone and the uniformly-rotating radiative interior beneath. \citet{SpiegelZahn92}  proposed the first hydrodynamical model, which here we call
SZ92, arguing that the tachocline is essentially in a steady state of thermal-wind balance, angular-momentum balance, and thermal equilibrium. Angular momentum transport in their model is assumed to be dominated by strongly anisotropic turbulence,  primarily horizontal owing to the strong stable stratification of the radiative interior.
By contrast, the heat transport is assumed to be dominated by a predominantly vertical diffusive heat flux owing to the thinness of the tachocline. In this paper, we demonstrate that these assumptions are not consistent with the new  model of stratified turbulence recently proposed by \citet{Chinietal22} and \citet{Shahal24}, which has been numerically validated by \citet{Garaudal24}. We then propose a simple self-consistent alternative to the SZ92 model, namely, a scenario wherein  angular momentum and heat transport are both dominated by horizontal turbulent diffusion. The thickness of the tachocline in the new model scales as $\Omega_\odot / N_m$, where $\Omega_\odot$ is the mean angular velocity of the Sun, and $N_m$ the buoyancy frequency  in the tachocline region. 
We discuss other properties of the model, and show that it has several desirable features, but does not resolve some of the other well-known problems of the SZ92 model.  
\end{abstract}

\section{Introduction}
\label{sec:intro}

The solar tachocline, whose existence was predicted by \citet{Spiegel72}, was unambiguously detected by helioseismology in the late 1980s \citep{JCDSchou88,Brownal89}. Across a layer whose thickness is at most a few per cent of the solar radius, the strong latitudinal shear that characterizes the rotation profile of the convection zone decays abruptly with depth, leaving the underlying radiative zone in a state of almost uniform rotation (at least as far down as p-mode seismology can discern). Why this is the case remains one of the dynamical mysteries of the solar interior, even though many possible explanations for the tachocline have been explored in the past three decades \citep[see the reviews in the book by][]{Tachocline2007}. This problem must be solved, however, because tachocline flows are known to play a central role in the transport of some chemical species between the radiative interior and the solar surface, and may also play a central role in generating the solar dynamo.  

\subsection{The Spiegel \& Zahn model of the tachocline}

 \citet{SpiegelZahn92} (SZ92 hereafter) were the first to study the dynamics of the tachocline, and demonstrated that it cannot remain thin 
 over long timescales in a purely laminar hydrodynamical model (in what follows, we describe as hydrodynamical any flow where the Lorentz force can be neglected). Specifically, they argued that slow meridional flows naturally cause the convection-zone shear to spread into the radiative interior on a timescale shorter than the age of the Sun, and concluded that this spread can be stopped only if strong anisotropic turbulent  stresses significantly contribute to the angular momentum balance (at least in their hydrodynamic study).  They then noted that the tachocline is in fact likely to be turbulent, and that the turbulence ought to be highly anisotropic, with horizontal velocities that are much larger than vertical velocities, because of the  strong stable stratification. They showed that if angular momentum is transported much more rapidly in the horizontal direction than in the vertical direction, it is possible to obtain steady solutions for the tachocline dynamics in which the latitudinal differential rotation and the meridional flows are both limited to a thin boundary layer at the base of the convection zone of thickness $h$, where
\begin{equation}
    h \simeq \left(\frac{\kappa}{\nu_h}\right)^{1/4}  \left(\frac{\Omega_\odot}{N_m}\right)^{1/2} r_{cz},
    \label{eq:hsz}
\end{equation}
(see their equation 5.19) in which $\kappa$ is the microscopic thermal diffusivity (which includes both radiative and conductive contributions), $\nu_h$ the assumed turbulent horizontal momentum diffusivity, $\Omega_\odot$ the mean angular velocity of the radiative interior, $N_m$ the mean buoyancy frequency in the tachocline, and $r_{cz}$ the radius of the base of the convection zone.  

This formula can easily be tested against modern observations.
The thickness of the tachocline inferred from helioseismology is generally said to be a few  per cent of the convection-zone radius, with observational estimates ranging from         $0.01r_{cz}$ to $0.1r_{cz}$ depending on how it is measured \citep{Kosovichev96,AntiaBasu1998,Corbardal1998,Charbonneaual99,ElliottGough99,AntiaBasu2011}. If the horizontal turbulent viscosity $\nu_h$ is due to instabilities of the rotational shear \citep[see][]{Zahn92}, one may reasonably assume on dimensional grounds that $\nu_h = O(\ell^2_h S_h)$, where $\ell_h$ is the characteristic length scale of that shear flow (and equivalently the horizontal length scale of the dominant turbulent eddies), and $S_h$ is the amplitude of the shear. Using a plausible estimate of $\ell_h = O(10^{10})$cm (which is somewhat smaller than, but of the order of $r_{cz}$), and $S_h =\Delta \Omega$, where $\Delta \Omega \simeq 0.1 \Omega_\odot$ is the amplitude of the differential rotation in the tachocline, we find that $\nu_h = O(10^{13})$cm$^2$/s. Combining this with $\kappa \simeq 10^7$cm$^2$/s, $\Omega_\odot \simeq 3\times 10^{-6}$s$^{-1}$, and $N_m \simeq 8 \times 10^{-4}$s$^{-1}$ (see Table  \ref{tab:fiducialvalues}) yields a predicted tachocline thickness $h = O(10^{-3})r_{cz}$, which is an order of magnitude smaller than the aforementioned observed values. 

 \begin{table}[h!]
\begin{center}
\begin{tabular}{ccc}
   \hline 
   Name & Variable & Value  \\
   \hline 
   Radius of base of convection zone & $r_{cz}$ & $5 \times 10^{10}$cm \\  
   Mean gravity & $g_m$ & $5\times 10^4$cm s$^{-2}$ \\
   Mean temperature & $T_m$ & $2\times 10^6$ K \\
   Mean density & $\rho_m$ & 0.2g cm$^{-3}$ \\
   Mean buoyancy frequency & $N_m$ & $8\times 10^{-4}$s$^{-1}$ \\
   Microscopic kinematic viscosity & $\nu$ & $30$cm$^2$s$^{-1}$ \\ 
   Microscopic thermal diffusivity & $\kappa$ & $10^7$cm$^2$s$^{-1}$ \\
   \hline
   Mean rotation rate of the radiative zone& $\Omega_\odot$ & $3 \times 10^{-6}$s$^{-1}$ \\
   Differential rotation of convection zone  & $\Delta \Omega$ & $3 \times 10^{-7}$s$^{-1}$
   \\
   Estimated Froude number $S_h / N_m$ & $Fr$ &  $\lesssim  10^{-3}$
 \\
 Estimated Rossby number $S_h / 2 \Omega_\odot$ & $Ro$ & $\lesssim 0.05$ \\ 
 Estimated Reynolds number $S_h \ell_h^2 / \nu$ & $Re$ & $\lesssim 10^{12}$ \\
   Estimated P\'eclet number $S_h \ell_h^2 / \kappa$  & $Pe$ & $\lesssim 10^6$\\ 
 \hline
\end{tabular}
\end{center}
\caption{Fiducial values for the mean solar tachocline properties adopted in this work, from \citet{Gough07}. The differential rotation of the convection zone is taken to be approximately half the difference between the equator-to-pole differential rotation. To estimate $Fr$, $Ro$, $Pe$ and $Re$, we used equation (\ref{eq:FrPedef}) with a horizontal length scale $\ell_h = 10^{10}$cm, and $S_h = \Delta \Omega$ at the top of the tachocline (but note that $S_h$ decreases rapidly with depth, so all four dimensionless numbers decrease with depth as well).}
\label{tab:fiducialvalues}
\end{table}

To shed light on this mismatch,
we point out that helioseismic inversions of the rotation profile only place upper limits on $h$ \citep{Corbardal1998}, because the resolution is too poor to do otherwise, and that the actual tachocline could in fact be much thinner. However, additional information from observations of trace elements (such as Li and Be) at the solar surface, and from the sound-speed profile beneath, which is seismically much more highly resolved, indicate that 
the tachocline is mixed, by turbulent diffusion according to \citet{BTZ1999} down through  
$h \gtrsim 0.01 r_{cz}$ beneath the convection zone, and by large-scale 
meridional flow according to \citet{ElliottGough99}, to $h \simeq 0.01 r_{cz}$.  
Requiring $h \simeq 0.03r_{cz}$, 
for example, a compromise between the seismic mixing estimates 
and the angular-momentum-based estimates, yields $\nu_h = O(10^8)$cm$^2$/s according to equation \eqref{eq:hsz}, which is perhaps {\it too} small to be realistic given the likely presence of horizontal shear instabilities.

This and several other problems with the SZ92 model have been pointed out before \citep{GoughMcIntyre98}, leading the scientific community to seek alternative solutions to prevent the convection-zone shear from spreading into the radiative interior. Most of these solutions involve the presence of magnetic fields \citep[see, e.g.][and Section \ref{sec:disc} for a brief discussion]{RudigerKitchatinov97,CharbonneauMacGregor93,GoughMcIntyre98,FDP2001,BrunZahn06,Strugarek-etal11,AAal13,WoodBrummell2018,Matilsky-etal22,Matilsky-etal24}. In this work, however, we pursue a different path, and revisit the SZ92 model in the light of recent theoretical progress made to quantify turbulent transport in shear-driven stratified turbulence \citep{Chinietal22,Shahal24}, which we summarize below. 

\subsection{Quantifying transport by stratified turbulence in the solar tachocline}
 
\citet{Chinietal22} and \citet{Shahal24} 
used multiscale asymptotic analysis to 
study the properties of strongly stratified turbulence driven by large-scale horizontal shear flows, which is precisely the case in the solar tachocline. \citet{Shahal24} discovered the existence of several dynamical regimes depending on the characteristic horizontal shear length scale $\ell_h$ (which they assume is the same as the injection scale of the horizontal eddies), the local horizontal shear $S_h$, the local buoyancy frequency $N_m$, as well as the microscopic kinematic viscosity $\nu$ and  thermal diffusivity $\kappa$. Each  regime occupies a specific region of parameter space, best represented in terms of the dimensionless parameters 
\begin{equation}
    Fr \equiv \frac{S_h}{N_m} , Re  \equiv \frac{S_h \ell_h^2}{\nu}  \mbox{  and  }  
    Pe  \equiv Pr Re \equiv \frac{S_h \ell_h^2}{\kappa}, 
    \label{eq:FrPedef}
\end{equation}
which are called the Froude, Reynolds and P\'eclet numbers of the large-scale horizontal flow, respectively. The quantity $Pr = \nu/\kappa$ is the Prandtl number. Estimates for these parameters at the top of the tachocline are given in Table \ref{tab:fiducialvalues}, using $S_h = \Delta \Omega$ and $\ell_h = O(10^{10})$cm as before. Notably, $Pr \simeq 10^{-6}$ in the bulk of the tachocline.

\begin{figure}
\centerline{\includegraphics[width=\textwidth]{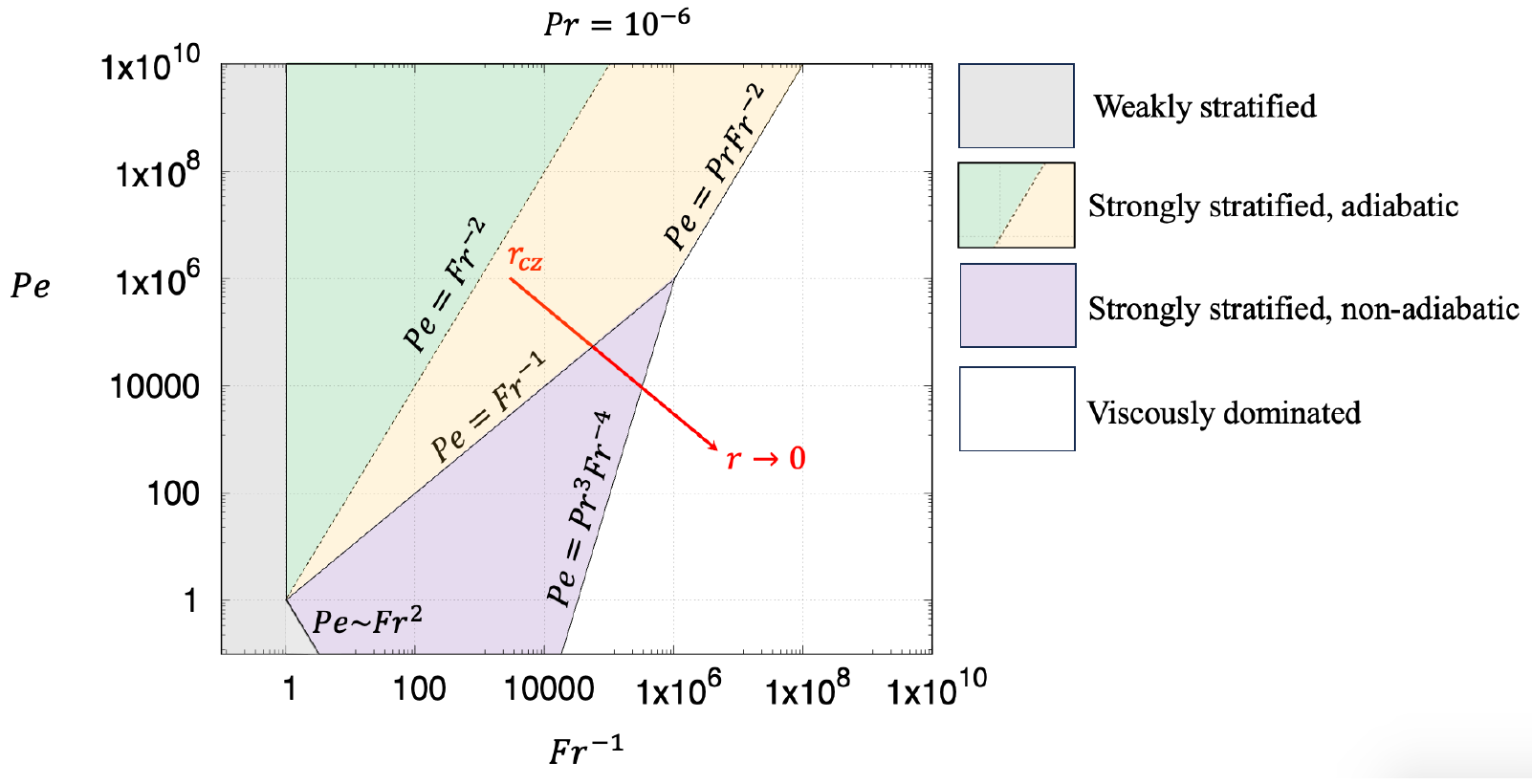}}
\caption{Regime diagram for stratified turbulence in the solar tachocline, which has a Prandtl number $Pr \simeq 10^{-6}$, based on \citet{Shahal24}. Regions of parameter space are colored according to the legend, see the main text for detail. The red arrow shows the path taken through parameter space going inwards from the top of the tachocline (at $r = r_{cz}$) where $Fr \simeq 4 \times 10^{-4}$ and $Pe = O(10^6)$ (see Table \ref{tab:fiducialvalues}). The path appears as a straight line because $Pe$ and $Fr$ are both proportional to $S_h$, which decreases as $r$ decreases, and we assume that $N_m$, $l_h$ and $\kappa$ are constant.}
\label{fig:regimes}
\end{figure}

Figure \ref{fig:regimes} \citep[adapted from][]{Shahal24} presents the partitioning of stratified turbulence parameter space for $Pr = 10^{-6}$. In the grey region, where the Froude number is generally large, the effects of stratification are negligible. In the green and yellow regions\footnote{The distinction between yellow and green regions is not relevant for the purpose of this work, but interested readers are invited to study the model of \citet{Shahal24} for more information.}, the turbulence is strongly anisotropic because of the large stratification ($Fr \ll 1$), and turbulent eddies transport heat adiabatically in both horizontal and vertical directions. This regime is delimited by $Fr = 1$ on the left, and by $Pe > {\rm max}(Fr^{-1},PrFr^{-2})$ from below. In the purple region, the turbulence is also strongly anisotropic, and the turbulent eddies transport heat adiabatically in the horizontal direction but not in the vertical direction. This non-adiabatic regime is  similar to the one described by \citet{Zahn92}. This region
is delimited by $Fr = 1$ from the left, and requires $Fr^{-1} > Pe > {\rm max}(Fr^2, Pr^3 Fr^{-4})$. Finally, the white region corresponds to parameters at which the turbulence is viscously suppressed.

The red arrow in Figure \ref{fig:regimes} shows, approximately, the values adopted by $Fr$ and $Pe$ on a path from the base of the convection zone inwards. We see that the Froude number is always very small on that path ($Fr^{-1} \gg 1$), consistent with the notion that turbulence is strongly influenced by stratification \citep{SpiegelZahn92}. We also see that the bulk of the tachocline near the base of the convection zone lies in the adiabatic regime of strongly stratified turbulence. Deeper down, the horizontal shear $S_h$ decreases, and $Pe$ and $Fr^{-1}$ both decrease in proportion. The turbulence briefly enters the non-adiabatic regime, and then rapidly becomes stabilized by viscosity. 

\citet{Chinietal22} and \citet{Shahal24} proposed simple scaling laws for the horizontal and vertical components of the turbulent viscosity ($\nu_h$ and $\nu_v$, respectively) and of the turbulent heat diffusivity ($\kappa_h$ and $\kappa_v$, respectively)
in each turbulent region of parameter space. They showed that these mixing coefficients scale as
\begin{eqnarray}
    \nu_h, \kappa_h \propto S_h \ell^2_h \mbox{ in all turbulent regimes,}  \label{eq:horizcoeffs}\\     
    \kappa_v \propto Fr^2 \kappa_h \mbox{ in all strongly stratified turbulent regimes, } \label{eq:kappavreg}\\
    \nu_v \propto Fr^{3/2} \nu_h \mbox{ in the strongly stratified adiabatic turbulent regime,} \label{eq:nondiffreg} \\ 
    \nu_v \propto (Fr^2/Pe)^{1/2} \nu_h \mbox{ in the strongly stratified non-adiabatic turbulent regime,} \label{eq:diffreg} 
\end{eqnarray}
where the coefficients of proportionality  are all of order unity. Crucially, these scalings have been validated by direct numerical experiments by \citet{Copeal20}, and \citet{Garaudal24}. 

For the sake of simplicity, we shall assume in what follows that the turbulence remains in the adiabatic regime throughout the tachocline, and unless stated otherwise, we use the scalings (\ref{eq:horizcoeffs}), (\ref{eq:kappavreg}) and (\ref{eq:nondiffreg}) to model the corresponding vertical and horizontal turbulent transport coefficients.  

\subsection{Revisiting SZ92}

In this work, we revisit the SZ92 model of the solar tachocline in the light of the newly available prescription for shear-driven stratified hydrodynamic turbulence, and closely examine each of its core assumptions in Section \ref{sec:cartmodel}. As we shall demonstrate, this exercise reveals a fundamental inconsistency in one of the central assumptions of the SZ92 model, namely their requirement that horizontal turbulent diffusion dominates the momentum transport, while vertical radiative diffusion dominates the heat transport. To correct the issue, we argue in Section \ref{sec:newmodel} that horizontal turbulent diffusion should dominate the transport of both momentum and heat. This yields a hydrodynamical model of the tachocline that is not only mathematically self-consistent, but also consistent with helioseismic observations of the tachocline thickness. Section \ref{sec:implications} discusses the implications of our findings for our understanding of the solar tachocline, and Section \ref{sec:disc} summarizes remaining problems with the model, and future work needed to solve them.

\section{Cartesian version of the SZ92 model}
\label{sec:cartmodel} 

\subsection{Model geometry, boundary conditions, and governing equations}
\label{sec:cartmodelderivation}

We model the tachocline using a Cartesian coordinate system, as do \citet{GaraudBrummell08} and \citet{GaraudAA09}, for instance. This enables us to obtain fully analytical, easily interpretable  solutions of the governing equations. Furthermore, we assume that gravity and the mean angular velocity are  constant in the tachocline, and, notwithstanding the spherical geometry 
of the Sun, are both aligned with the $z$ direction (${\bf g} = - g_m {\bf e}_z$ and ${\bf \Omega}_\odot= \Omega_\odot {\bf e}_z$; values for $g_m$ and $\Omega_\odot$ are given in Table \ref{tab:fiducialvalues}). Finally, we assume as in SZ92 that any magnetic field that may be present in the tachocline has a negligible effect on its dynamics, thus adopting a purely hydrodynamic approach. This assumption may not be valid in the Sun, as discussed in Section \ref{sec:disc}, but we nevertheless use it here for consistency with SZ92.

The base of the convection zone is assumed to be located at $z = 0$, negative $z$ being in the radiative zone below. The unit vector ${\bf e}_x$ points in the `azimuthal' direction and ${\bf e}_y$ points in the `latitudinal' direction. 
At the top of the tachocline (which we define to coincide with the base of the convection zone), we assume the presence of some large-scale `differential rotation', which is modeled as  the azimuthal shear flow
\begin{equation}
{\bf u}_{cz}(x,y,0,t) = u_{cz} \cos\left( \frac{y}{r_{cz}} \right){\bf e}_x, 
\label{eq:uczbc_cart}
\end{equation}
where $t$ is time and $r_{cz}$ is a characteristic
length 
scale of that flow, which would be the radius of the base of the convection zone in a spherical shell. The quantity $u_{cz}$ is analogous to the amplitude of the azimuthal velocity associated with the differential rotation near the top of the solar tachocline, and in this study any (long-term) temporal variation of $u_{cz}$ is ignored. Characteristic values are: $r_{cz} \simeq 5 \times 10^{10}$cm, and $u_{cz} \simeq  r_{cz} \Delta \Omega\simeq 10^4$cm/s, using the fact that $\Delta \Omega \simeq 0.1 \Omega_\odot \simeq 3 \times 10^{-7}$s$^{-1}$ at the top of the tachocline.

The radiative region in this model is assumed to be stably stratified with constant buoyancy frequency $N_m$. Because the tachocline is so thin compared to the local pressure scale height $H_p$ ($H_p\approx6\times10^9$ cm at the base of the solar convection zone), we can use the Boussinesq equations for a rotating stably stratified perfect gas, which are (e.g., \citealt{SpiegelVeronis1960}):
\begin{eqnarray}
\rho_m  \left( \frac{\partial \bu}{\partial t} + 2{\bf \Omega}_\odot\times \bu \right) = - \nabla p + \rho {\bf g} + \nabla \cdot {\bf \Pi}, \label{eq:mom1} \\
\frac{\rho}{\rho_m} = - \frac{T}{T_m}, \label{eq:eos}\\ 
\frac{\partial T}{\partial t }  + \frac{N_m^2 T_m}{ g} w =  \nabla \cdot {\bf F}_T, \label{eq:temp1} \\
\bnabla \cdot \bu = 0,
\label{eq:goveqs_cart}
\end{eqnarray}
where ${\bf u} = (u,v,w)$ represents the {\it large-scale} azimuthal and meridional flow, $\rho_m$ and $T_m$ are the mean density and temperature of the tachocline (respectively), and $p$, $T$, and $\rho$ are the pressure, temperature, and density perturbations away from hydrostatic equilibrium caused by the presence of the flow. All of the dependent variables are assumed to be independent of $x$ (`axially symmetric'), and vary only on large length and time scales compared with the turbulent fluctuations. The quantities ${\bf \Pi}$ and ${\bf F}_T$ represent the momentum stress tensor  and the thermal flux, respectively, and each contain the sum of microscopic diffusive contributions and macroscopic turbulent contributions arising from the transport of momentum and temperature by fast, small-scale, horizontally isotropic  eddies with greatly differing vertical and horizontal velocities. In these equations, we have neglected the centrifugal force and solar oblateness, for consistency with SZ92. We have also neglected nonlinear terms involving the large-scale flows in the momentum and temperature equations, as in SZ92, because the latter are thought to be small. This can be verified {\it a posteriori} (see section \ref{sec:selfconst}). 

We now continue to make the same series of assumptions as in SZ92. First, we assume that the dynamics continues to be governed by hydrostatic equilibrium for the perturbations $p$, $\rho$, and $T$, so the vertical component of the momentum equation reduces to
\begin{equation}
\frac{\partial p}{\partial z} = - \rho g_m =  \frac{\rho_m}{T_m} T g_m, 
\label{eq:hydrostat_cart}
\end{equation}
where the second equality is obtained using the linearized equation of state (\ref{eq:eos}).
Second, we assume that the dynamics is governed by heliostrophic (the analogue of geostrophic) equilibrium, so the latitudinal component of the momentum equation reduces to a balance between the horizontal pressure gradient and the Coriolis force:
\begin{equation}
2 \rho_m \Omega_\odot u = - \frac{\partial p}{\partial y} .
\label{eq:geostroph_cart}
\end{equation} 
Eliminating pressure between equations (\ref{eq:hydrostat_cart}) and (\ref{eq:geostroph_cart}), we obtain the usual thermal-wind balance equation
\begin{equation}
    2 \Omega_\odot \frac{\partial u}{\partial z} = -\frac{g_m}{T_m} \frac{\partial T}{\partial y} .   
    \label{eq:tw_cart}
    \end{equation}
The incompressibility condition, assuming axial symmetry, implies
\begin{equation}
    \frac{\partial v}{\partial y} + \frac{\partial w}{\partial z}  = 0 ,
\label{eq:incomp_cart}
\end{equation}
permitting the introduction of a stream function 
$\psi$ satisfying 
\begin{equation}
\frac{\partial \psi}{\partial z }= -v,\;\; \frac{\partial \psi}{\partial y} = w .
\label{eq:streamfn}
\end{equation}
Finally, in the temperature equation and azimuthal component of the momentum equation,  we assume that the divergence of the fluxes of temperature and momentum can be written in terms 
of diagonal diffusion tensors $\tilde{\kappa}$ and $\tilde{\nu}$ as
\begin{eqnarray}
\nabla \cdot {\bf F}_T = \frac{\partial }{\partial z } \left( \tilde{\kappa}_v \frac{\partial T}{\partial z} \right)+ \tilde{\kappa}_h \frac{\partial^2 T}{\partial y^2} ,
\label{eq:fluxes_cart} \\ (\nabla \cdot {\bf \Pi})_x =  \rho_m \frac{\partial }{\partial z } \left( \tilde{\nu}_v\frac{\partial u}{\partial z}\right) + \rho_m \tilde{\nu}_h\frac{\partial^2 u}{\partial y^2} , \label{eq:rscart}  
\end{eqnarray}
where $\tilde{\nu}_h$ is the sum of the microscopic kinematic viscosity $\nu$ and the component $\nu_h$ of the turbulent viscosity tensor arising from the horizontal motion of small-scale eddies, and $\tilde{\kappa}_h$ is similarly the microscopic thermal diffusivity $\kappa$ plus the horizontal component $\kappa_h$ of the turbulent diffusivity, and  $\tilde{\nu}_v$ and $\tilde{\kappa}_v$ are their corresponding vertical counterparts. As $\nu$ and $\kappa$ are determined by the properties of the background temperature and density, they are assumed to be constant in the tachocline (which is very thin). The turbulent coefficients ($\nu_h$, $\kappa_h$, $\nu_v$ and $\kappa_v$), by contrast, can depend on depth through their dependence on the local horizontal shear, but are assumed for simplicity to be independent of $x$ or $y$ in this simple model. 

One may rightfully question whether momentum and temperature fluxes in strongly anisotropic rotating stratified turbulence can be modeled in this manner, namely taking the form of anisotropic turbulent diffusion. In particular, it has been suggested that layerwise two-dimensional turbulent transport tends to diffuse vorticity rather than momentum \citep[e.g.][]{1935RSPSA.151..494T,1968JFM....32..437G}; indeed  strictly two-dimensional turbulence is known to have what has been called anti-diffusive, rather than diffusive,  properties with regard to momentum transport \citep{Rhines1975,VallisMaltrud1993,Tobiasal2007}.  Moreover, there is a possibility that a residual of that property is retained when the turbulence is three-dimensional but strongly anisotropic. Nevertheless, following SZ92, we proceed by adopting this simple model for now, and look for axisymmetric solutions of the governing equations. 

Using (\ref{eq:rscart}) and (\ref{eq:fluxes_cart}), and the definition of the stream function, the azimuthal component of the momentum equation \eqref{eq:mom1} reduces to: 
\begin{equation}
\frac{\partial u}{\partial t} + 2  \Omega_\odot \frac{\partial \psi}{\partial z}  = \frac{\partial }{\partial z } \left( \tilde\nu_v \frac{\partial u}{\partial z} \right)+ \tilde{\nu}_h  \frac{\partial^2 u}{\partial y^2 },  \label{eq:azimuth_cart}
\end{equation}
and similarly the temperature equation becomes 
\begin{equation}
 \frac{\partial T}{\partial t} +  \frac{N_m^2 T_m}{ g_m}\frac{\partial \psi}{\partial y} = \frac{\partial }{\partial z } \left(  \tilde{\kappa}_v \frac{\partial T}{\partial z} \right)+ \tilde{\kappa}_h  \frac{\partial^2 T}{\partial y^2 } .\label{eq:temp_cart} 
\end{equation}
Eliminating $\psi$ between equations (\ref{eq:azimuth_cart}) and (\ref{eq:temp_cart}) and substituting for $T$ with equation (\ref{eq:tw_cart}) yields the following master equation:
\begin{equation}
\frac{\partial^2}{\partial y^2}\left(\frac{\partial }{\partial t}-\tilde{\nu}_h\frac{\partial^2 }{\partial y^2}- \frac{\partial }{\partial z}\tilde{\nu}_v\frac{\partial }{\partial z}\right)u +\frac{4 \Omega_\odot^2}{N_m^2}\frac{\partial}{\partial z}\left(\frac{\partial }{\partial t}-\tilde{\kappa}_h\frac{\partial^2 }{\partial y^2}-\frac{\partial }{\partial z}\tilde{\kappa}_v \frac{\partial }{\partial z} \right)\frac{\partial u}{\partial z}=0 .\label{eq: master}
\end{equation} 

 For pedagogical purposes, we now reproduce the SZ92 solution, here in Cartesian coordinates, and critically assess its validity in the light of the turbulence model introduced in Section \ref{sec:intro}. In the following section we contrast it with an alternative. 

\subsection{The turbulent tachocline of SZ92}
\label{sec:zahncart}
 
Arguing that stratified turbulence is strongly anisotropic, SZ92 kept only the horizontal contribution 
$\tilde{\nu}_h \partial^2 u / \partial y^2$ to the viscous stress in equation (\ref{eq:azimuth_cart}). Noting that the tachocline is very thin,  they kept only the vertical contribution $\tilde{\kappa}_v \partial^2 T / \partial z^2$ to the temperature flux in equation (\ref{eq:temp_cart}). These assumptions are self-consistent provided the tachocline thickness $h$ satisfies
\begin{equation}
    \frac{\tilde{\nu}_v}{\tilde{\nu}_h} \ll \frac{h^2}{r_{cz}^2} \ll \frac{\tilde{\kappa}_v}{\tilde{\kappa}_h}.
    \label{eq:selfconSZ92}
\end{equation} 
Finally, they assumed that the temperature fluctuation $T$ rapidly reaches a state of thermal equilibrium to justify dropping the time derivative in equation (\ref{eq:temp_cart}).  These simplifications result in a simplified version of the master equation (\ref{eq: master}) in which 
the term in $\tilde \nu_v$ is absent from 
the first bracketed term, and the time derivative and the term in $\tilde\kappa_h$ are both absent from the second bracketed term:
\begin{equation}
\frac{\partial^2}{\partial y^2}\left(\frac{\partial }{\partial t}-\tilde{\nu}_h\frac{\partial^2 }{\partial y^2}\right)u = \frac{4 \Omega_\odot^2}{N_m^2}\frac{\partial^2}{\partial z^2}\left(\tilde{\kappa}_v \frac{\partial^2 u}{\partial z^2} \right)  .\label{eq: masterSZ} 
\end{equation}

It is easy to verify that  solutions for $u$, $v$, $w$ and $T$, given the boundary condition (\ref{eq:uczbc_cart}), can be written  in the separated form:
$u(y,z,t) = \hat u(z,t) \cos(y/r_{cz})$, $v(y,z,t) = \hat v(z,t) \cos(y/r_{cz})$,  $w(y,z,t) = \hat w(z,t) \sin(y/r_{cz})$, and $T(y,z,t) = \hat T(z,t) \sin(y/r_{cz})$. On substituting the ansatz for $u$ 
into (\ref{eq: masterSZ})  one obtains 
\begin{equation}
\frac{\partial \hat u}{\partial t} + 4   r^2_{cz} \frac{ \Omega^2_\odot}{N_m^2 }  \frac{\partial^2 }{\partial z^2 } \left( \tilde{\kappa}_v \frac{\partial^2 \hat u}{\partial z^2} \right)   = - \frac{ \tilde{\nu}_h}{r_{cz}^2}  \hat u .  
\label{eq:azimuth_cart2}
\end{equation}
In this limit the effect of the meridional flow thus takes the form of hyperdiffusion (second term on the left-hand side). As noted by SZ92, an unbalanced hyperdiffusion would cause the tachocline to grow quite quickly, even if it were infinitesimally thin at early times. However, in the presence of rapid horizontal turbulent momentum diffusion ($\tilde{\nu}_h$), $\hat u$ can eventually attain a steady state with a thin tachocline. Assuming  the diffusion coefficients (microscopic and turbulent) to be constant in the tachocline, as in SZ92, the steady-state equation reads:
\begin{equation}
\frac{\partial^4 \hat u}{\partial z^4} = - \frac{\tilde\nu_h}{\tilde{\kappa}_v} \frac{N_m^2}{4\Omega_\odot^2}  \frac{\hat u}{r_{cz}^4}. 
\label{eq:SZ92PDE_cart}
\end{equation}
This equation is essentially the same as (5.15) in SZ92, aside from an order-unity coefficient that arises from the use of a simplified Cartesian rather than spherical geometry. Seeking solutions of the form $\hat u \propto e^{k_z z}$ (with the real part of $k_z$ positive to ensure that $\hat u$  decays as $z \rightarrow -\infty$), we have
\begin{equation}
k_z^4 = - \frac{\tilde{\nu}_h}{\tilde{\kappa}_v} \frac{N_m^2}{4\Omega_\odot^2} r_{cz}^{-4}, \mbox{ so } k_z = \left(\frac{1}{2} \pm \frac{i}{2} \right) \left(\frac{\tilde{\nu}_h}{\tilde{\kappa}_v}\right)^{1/4}  \left(\frac{N_m}{\Omega_\odot}\right)^{1/2} r_{cz}^{-1}. 
\end{equation}
We see that physical solutions decay exponentially with depth below the convection zone, on a length scale  given by the inverse of the real part of $k_z$,
namely 
\begin{equation}
    h_{\rm SZ} = 2 \left(\frac{\tilde{\kappa}_v}{\tilde{\nu}_h}\right)^{1/4}  \left(\frac{\Omega_\odot}{N_m}\right)^{1/2} r_{cz},
    \label{eq:hsz_cart}
\end{equation}
which is the Cartesian  equivalent of equation (5.19) of SZ92. Using the values of $\Omega_\odot$ and $N_m$ listed in Table \ref{tab:fiducialvalues}, we obtain
\begin{equation}
    h_{\rm SZ} \simeq 0.1 \left(\frac{\tilde{\kappa}_v}{\tilde{\nu}_h}\right)^{1/4}r_{cz}.
\end{equation}
Note that there is little point in keeping more than one significant digit in the solution because the approximation made in using a Cartesian geometry  unavoidably introduces $O(1)$ corrections. The remaining diffusivity ratio depends on the turbulent diffusivities $\nu_h$ and $\kappa_v$, which are not known {\it a priori}. Using the estimates given in equations (\ref{eq:horizcoeffs}) and (\ref{eq:kappavreg}),
 \begin{equation}
     \nu_h, \kappa_h = O( S_h \ell_h^2 ),\;\; \kappa_v = O(S_h^2 / N_m^2) \kappa_h ,
     \label{eq:nuh_cart}
 \end{equation}
taking $S_h \simeq \Delta \Omega$ at the top of the tachocline, and $\ell_h = O(10^{10})$cm as before, we find $\nu_h, \kappa_h =  O(10^{13})$cm$^2$/s, which is much larger than their microscopic counterparts. We also have $\kappa_v = O( 10^{-7}) \kappa_h 
= O( 10^{6})$cm$^2$/s $<  \kappa$ which confirms that the vertical turbulent diffusion of heat can probably be neglected  in this  tachocline model, so $\tilde{\kappa}_v \simeq \kappa$ as originally assumed by SZ92. We finally obtain \begin{equation}
  h_{\rm SZ} \simeq 0.1  \left(\frac{\kappa}{\tilde{\nu}_h}\right)^{1/4}r_{cz} \sim O(10^{-3}) r_{cz}, 
\label{eq:hsz_derived}
\end{equation} 
which recovers the estimate provided in Section \ref{sec:intro}. 

Knowing the solution for  $u$, we can find the solution for the latitudinal component of the meridional flow within the tachocline. Indeed, from the steady-state limit of the azimuthal momentum equation (\ref{eq:azimuth_cart}), 
\begin{equation}
v \simeq - \frac{\nu_h}{2\Omega_\odot} \frac{\partial^2 u}{\partial y^2} = \frac{\nu_h}{2\Omega_\odot} \frac{u}{r_{cz}^2}  ,
\label{eq:v_eq_cart}
\end{equation}
we see that $v$ also decays
 with depth on the same length scale as $u$. At the top of the tachocline where $u \simeq u_{cz} \simeq r_{cz} \Delta \Omega$,  a typical value of $v$ is given by  
 \begin{equation}
     v \sim  \frac{\nu_h}{2\Omega_\odot} \frac{\Delta \Omega}{r_{cz}}   \sim O(10)\ {\rm cm/s}.
     \label{eq:v_est_cart}
 \end{equation}
  Finally, we use this result with the continuity equation to estimate the characteristic vertical velocity of meridional flows at the top of the tachocline to be: 
 \begin{equation}
     w \sim \frac{h_{\rm SZ}}{r_{cz}} v \sim O(10^{-2}) \ {\rm cm/s}.
 \end{equation}
  Both $v$ and $w$ then decrease exponentially with depth beneath the convection zone on the length scale $h_{\rm SZ}$. 

\subsection{Failure of the model}\label{sec:SZfailure}

Self-consistency of the SZ92 model requires both inequalities in (\ref{eq:selfconSZ92}) to hold. The first is needed in order to neglect the vertical turbulent diffusion of momentum in (\ref{eq:azimuth_cart}), while the second is needed to neglect the horizontal turbulent diffusion of heat in (\ref{eq:temp_cart}).  
We estimated in the previous section that $\tilde{\kappa}_v \simeq \kappa \simeq 10^7$cm$^2$/s and $\tilde{\kappa}_h  = O(10^{13})$cm$^2$/s, so $\tilde{\kappa}_v/\tilde{\kappa}_h = O(10^{-6})$ which is of the same order or smaller than $h_{\rm SZ}^2 / r_{cz}^2$, hereby invalidating the second inequality in  (\ref{eq:selfconSZ92}).
In other words, we have demonstrated that
one or more of the core assumptions of the SZ92 model is not consistent with the others: the transport of heat in the SZ92  tachocline solution cannot be dominated by vertical diffusion as assumed, but instead, is probably dominated by  horizontal turbulent diffusion.

\section{An alternative model of the turbulent tachocline}
\label{sec:newmodel}

\citet{SpiegelZahn92} recognized that a different tachocline model would be needed if (\ref{eq:selfconSZ92}) does not hold.
An alternative, and perhaps more natural, approach is to retain the horizontal turbulent diffusion term  and neglect the vertical one in equation (\ref{eq:temp_cart}). 
Such a procedure is self-consistent if  
\begin{equation}
  \frac{\tilde{\kappa}_v} {\tilde{\kappa}_h },\;    \frac{\tilde{\nu}_v}{\tilde{\nu}_h} \ll  \frac{h^2}{r_{cz}^2}.
    \label{eq:selfconnew}
\end{equation}
Furthermore, we show in the Appendix that the time derivative in  (\ref{eq:temp_cart}) cannot be ignored in this limit.  The master equation (\ref{eq: master}) becomes
\begin{equation}
 \frac{\partial} {\partial t}\left(  \frac{\partial^2 \hat u}{\partial z^2}- \frac{N_m^2}{ 4 \Omega_\odot^2   } \frac{ \hat u }{r^2_{cz}} \right) = \frac{N_m^2}{ 4 \Omega_\odot^2  } \frac{\tilde{\nu}_h}{r_{cz}^2} \frac{\hat u}{r_{cz}^2}   - 
 \frac{\partial}{\partial z}\left( \frac{\tilde{\kappa}_h}{r_{cz}^2}    \frac{\partial \hat u}{\partial z} \right) .
\label{eq:newPDE_cart}
\end{equation}  
 This equation is quite different from the one obtained in the SZ92 model: notably, the hyperdiffusion term has disappeared, and $\hat u$ evolves towards a steady state satisfying 
\begin{equation}
 \frac{\partial}{\partial z}\left[ \tilde{\kappa}_h   \frac{\partial \hat u}{\partial z} \right]=  \frac{N_m^2}{ 4 \Omega_\odot^2  } \frac{\tilde{\nu}_h}{r_{cz}^2} \hat u   .
\label{eq:newPDE_steady}
\end{equation} 

To gain  insight into the temporal evolution of this new tachocline model, it is useful first to explore what happens when $\nu_h$ and $\kappa_h$ are both constant. For simplicity we take them to be equal to each other, and much larger than their microscopic counterparts. In reality, $\nu_h$ and $\kappa_h$ are indeed expected to be approximately equal, but also to vary rapidly with depth. We study in Section \ref{sec:varynucart} the steady-state solution of (\ref{eq:newPDE_steady}) with spatially variable turbulent diffusivities.

\subsection{Case of constant and equal $\kappa_h,\nu_h$}
\label{sec:constnucart}

In this section only, we assume for simplicity that $\kappa_h = \nu_h \equiv D_h$ are  constant, and that $D_h \gg \nu,\kappa$. In that case, equation (\ref{eq:newPDE_cart}) reduces to 
\begin{equation}
 \frac{\partial} {\partial t}\left(  \frac{\partial^2 \hat u}{\partial z^2}- \frac{N_m^2}{ 4 \Omega_\odot^2   } \frac{ \hat u }{r^2_{cz}} \right) = - \frac{D_h}{r_{cz}^2} \left(  \frac{\partial^2 \hat u}{\partial z^2}- \frac{N_m^2}{ 4 \Omega_\odot^2   } \frac{ \hat u }{r^2_{cz}} \right) .
\label{eq:newPDE_cart2}
\end{equation} 
This indicates that $\hat u$ converges exponentially fast to its steady state, on the short characteristic timescale $\tau_h = r_{cz}^2 / D_h$, i.e. the horizontal turbulent diffusion timescale (from Table \ref{tab:fiducialvalues}, $\tau_h  = O(10)$ yr for the solar tachocline assuming $D_h = O(10^{13})$ cm$^2/$s).
The steady state for $\hat u$ satisfies a simple equation whose solutions are exponential in  $z$, such that 
\begin{equation}
 \hat u(z)  = u_{cz} \exp\left( \frac{z}{h} \right)  \mbox{  for  } z < 0, 
 \label{eq:cart_sol2}
\end{equation}
where again $h$ is the characteristic tachocline thickness, now given by 
\begin{equation}
h = 2 
\frac{\Omega_\odot }{N_m } r_{cz}.
\label{eq:newh_cart}
\end{equation}
As already stated in SZ92, in that case {\it ``the tachocline thickness reduces to about the scaleheight of the adiabatic adjustment layer in the present-day Sun"}.
Adopting the characteristic values listed in Table \ref{tab:fiducialvalues}, we obtain 
\begin{equation}
h = O( 10^{-2}) r_{cz},
\end{equation}
which is somewhat larger than $h_{\rm SZ}$. 

The estimated value of the latitudinal flow velocity $v$ in this model is the same as in the SZ92 case, namely of the order of 10cm/s, since it is given by the same equation (\ref{eq:v_est_cart}). The vertical flow velocity is given by 
\begin{equation}
    w \sim \frac{h}{r_{cz}} v = O(0.1)  \mbox{cm/s}.
\end{equation} 

Crucially, we find that this time the model is self-consistent. Earlier, we had shown that $\tilde{\kappa}_v/ \tilde{\kappa}_h =  O(10^{-6})$. We can estimate $\nu_v$ in the adiabatic turbulent regime from (\ref{eq:nondiffreg}) to be $\nu_v \sim Fr^{3/2} \nu_h = O(  10^{-5}) \nu_h \gg \nu$, so $\tilde{\nu}_v / \tilde{\nu}_h = O(10^{-5}$).  Finally, from (\ref{eq:newh_cart}) we have 
$h^2 / r_{cz}^2 = O(  10^{-4})$, which is indeed significantly larger than both $\tilde{\nu}_v / \tilde{\nu}_h$ and  $\tilde{\kappa}_v/ \tilde{\kappa}_h$ as required.  

Encouraged by these results, we now turn to the more realistic case of non-constant transport coefficients, and this time focus on the steady-state structure of the tachocline because the transient evolution is so rapid.

\subsection{$\kappa_h$, $\nu_h$ proportional to the amplitude of the shear}
\label{sec:varynucart}

From equation (\ref{eq:horizcoeffs}) we expect $\nu_h$ and $\kappa_h$  to vary rapidly with $z$ in the tachocline: if the turbulence is driven by the horizontal shear, then the transport coefficients ought to depend on the horizontal shearing rate $S_h$, which in turn decays with depth in step with the decay of $\hat u$. We now take this nonlinear effect into account, and 
seek steady  solutions of (\ref{eq:newPDE_cart}). Consistent with the turbulence model presented in Section \ref{sec:intro}, we posit that $\kappa_h$ and $\nu_h$ are both proportional to the local horizontal flow amplitude $|\hat u(z)|$, as in (\ref{eq:horizcoeffs}):
\begin{equation}
\kappa_h(z) = C_{\kappa} |\hat u(z)| \ell_h \mbox{ and }  \nu_h(z) = C_{\nu} |\hat u(z)| \ell_h,
\end{equation}
where $C_\kappa$ and $C_\nu$ are two constants of order unity, and $\ell_h$ is the horizontal extent of the largest eddies (which is assumed to be constant). We neglect the microscopic diffusivities $\nu$ and $\kappa$ because $\nu_h \gg \nu$ and $\kappa_h \gg \kappa$.

With these assumptions, the steady state equation (\ref{eq:newPDE_steady}) becomes 
\begin{equation}
\frac{d}{d z} \left[ C_\kappa |\hat u|  \frac{d \hat u}{d z}   \right]  =  \frac{N_m^2}{4\Omega_\odot^2}  \frac{C_\nu |\hat u|  \hat u }{r_{cz}^2} ,
\end{equation}
which simplifies to
\begin{equation}
\frac{d^2 \hat u^2}{d z^2}   = \frac{C_\nu}{C_\kappa} \frac{ N_m^2 }{2\Omega_\odot^2 }\frac{\hat u^2}{r_{cz}^2} , 
\end{equation}
whether $\hat u$ is positive or negative. Physical solutions   decay exponentially with depth as before, and are given by 
\begin{equation}
\hat u(z) = u_{cz} \exp\left( \frac{z}{h}\right),
\end{equation}
where this time
\begin{equation}
h =2 \sqrt{2}   \sqrt{\frac{C_\kappa}{ C_\nu}} \frac{\Omega_\odot }{N_m }r_{cz}  =  2 \sqrt{2}   \sqrt{\frac{\kappa_h}{ \nu_h}} \frac{\Omega_\odot }{N_m }r_{cz}. 
\label{eq:hcart}
\end{equation}
 We therefore recover almost the same formula as in the constant-viscosity/thermal-diffusivity case (equation \ref{eq:newh_cart}), except for the order-unity factor $\sqrt{2 C_\kappa/C_\nu}$. As a result, the predicted thickness of the tachocline has the same scaling with the model parameters as before. 
With regard to the meridional circulation, equation \eqref{eq:v_eq_cart} now yields 
\begin{equation}\label{eq:v_eq_cart2}
\hat v(z) =  \frac{\nu_h}{2\Omega_\odot} \frac{\hat u}{r_{cz}^2} = \frac{C_\nu |\hat u| \ell_h}{2\Omega_\odot } \frac{\hat u}{r^2_{cz}} \propto \hat u^2. 
\end{equation}
This demonstrates that $\hat v(z)$ decays with depth on the same lengthscale as $\hat u^2(z)$, namely $h/2$. 
The same statement applies to the vertical flow velocity $\hat w$.

\subsection{Self-consistency check.}
\label{sec:selfconst}

Using the formulae derived for $\hat u$ and $\hat v$, we now establish the conditions under which the assumptions made in deriving the original set of equations apply. The ratio of the neglected horizontal advection terms $\bu \cdot \nabla u$ and $\bu \cdot \nabla T$ to the horizontal turbulent diffusion terms $\nu_h \partial^2 u/\partial y^2$ or $\kappa_h \partial^2 T/\partial y^2$ kept in both the momentum and temperature equations (\ref{eq:mom1}) and (\ref{eq:temp1}) is  \begin{equation}
\frac{ \hat v r_{cz} }{\nu_h}   = \frac{\hat u}{2\Omega_\odot r_{cz}} 
\end{equation}
using equation \eqref{eq:v_eq_cart2}. This term is largest at the top of the tachocline, where $\hat u \simeq \hat u_{cz} = r_{cz} \Delta \Omega$, and thus $\hat{u}/(2\Omega_\odot r_{cz}) \le \Delta \Omega / 2 \Omega_\odot \equiv Ro$. This shows that the nonlinear advection terms can indeed be neglected in that region provided the Rossby number $Ro$ is small, which is the case in the tachocline (see Table \ref{tab:fiducialvalues}).

The next step is to show that the tachocline is indeed in thermal-wind balance as assumed. This is verified in the Appendix. Finally
we also need to be able to neglect the vertical transport terms in comparison with the horizontal transport terms for this model to be self-consistent, which requires (\ref{eq:selfconnew}) to hold. In Section \ref{sec:SZfailure} we noted that $\tilde{\kappa}_v/\tilde{\kappa}_h = O(10^{-6})$. In the adiabatic regime, $\nu_v = Fr^{3/2} \nu_h$ (see equation (\ref{eq:nondiffreg})), so $\tilde{\nu}_v / \tilde{\nu}_h = O(10^{-5}) > \tilde{\kappa}_v / \tilde{\kappa}_h$. In other words, we need only to verify that the inequality (\ref{eq:selfconnew}) holds for the turbulent viscosities. Using the estimated value of $h$ given in (\ref{eq:hcart}) and recalling that $Fr= S_h/N_m$, 
we find that the vertical diffusion terms can be neglected provided
\begin{equation}
 \left( \frac{S_h}{N_m } \right)^{3/2}  \ll 8\left(\frac{\kappa_h}{\nu_h}\right)\frac{\Omega_\odot^2}{N_m^2} \implies \frac{S_h}{\Omega_\odot}  \ll  4\left(\frac{\kappa_h}{\nu_h}\right)^{2/3}\left( \frac{\Omega_\odot}{N_m}  \right)^{1/3}.
 \label{eq:selfconst_main}
\end{equation}
As long as  $\kappa_h \simeq \nu_h$, this inequality holds at the top of the tachocline where $S_h \simeq \Delta \Omega$, and 
$4(\Omega_\odot/N_m)^{1/3}\simeq 0.6$ while $\Delta \Omega/ \Omega_\odot \simeq 0.1$. The inequality also holds throughout the tachocline, because $S_h$ decreases with depth much faster than $N_m^{1/3}$ increases.

\section{Summary and Implications}
\label{sec:implications}

SZ92 assumed that the tachocline dynamics is primarily controlled by hydrodynamic processes that include advection by large-scale meridional flows, and turbulent diffusion by strongly anisotropic stratified turbulence. Under the same assumptions, and using the recently derived model for stratified turbulence proposed by \citet{Chinietal22} and \citet{Shahal24}, 
we have demonstrated that the tachocline model of SZ92 is not self-consistent. 
In particular, we have shown that the horizontal turbulent diffusion of heat cannot be neglected as had originally been assumed, and instead plays a dominant role in establishing  thermal equilibrium in the tachocline.

We then proposed and studied an alternative model in which the turbulent anisotropic thermal diffusion is predominantly horizontal, and horizontally isotropic. The new model makes the following predictions: 
\begin{itemize}    
    \item In a steady-state, the latitudinal shear decays with depth exponentially away from the base of the convection zone, on a characteristic length scale  
\begin{equation}
h = 2\sqrt{2} \sqrt{\frac{\kappa_h}{\nu_h}} \frac{\Omega_\odot}{N_m} r_{cz},
\label{eq:hfinal}
\end{equation}
which we define to be the tachocline thickness. 
\item In a steady state, the meridional flows also decay with depth exponentially, but on a length scale that is half $h$. Near the top of the tachocline, the typical latitudinal and vertical flow velocities are 
\begin{equation}
    v \sim \frac{1}{2} \frac{\Delta \Omega}{\Omega_\odot} \frac{\nu_h}{ r_{cz}}, \quad \quad  
    w \sim \frac{h}{r_{cz}} v. 
\end{equation}
\item The tachocline in the new model relaxes towards a steady state on the fast horizontal turbulent diffusion time scale $\tau_h \sim r_{cz}^2/D_h = O(10)$ yr.  
\item The model is self consistent as long as 
\begin{equation}
    \frac{\Delta \Omega  }{\Omega_\odot}  \ll 4 \left( \frac{\Omega_\odot}{N_m}  \right)^{1/3},
\end{equation}
which is indeed satisfied in the present-day solar tachocline. 
\end{itemize}
Using the typical values listed in Table \ref{tab:fiducialvalues} for properties of the solar interior today, and assuming that $\kappa_h \simeq \nu_h$, we find that $h = O(10^{-2} r_{cz})$, which is consistent with helioseismic observations \citep{ElliottGough99,Charbonneaual99}.  The latitudinal flow velocities  at the base of the convection zone would be $v = O(10)$cm/s, and the vertical flow velocities would be 
$w = O(0.1)$cm/s, resulting in a ventilation time of  $O(100)$yr, as in the SZ92 model. Earlier in the life of the Sun, when the rotation rate was much larger, the tachocline would have been correspondingly thicker.  

\section{Discussion}
\label{sec:disc}

We have presented a simple self-consistent hydrodynamic  model representing a turbulent tachocline that assumes the turbulence to act as an anisotropic diffusion, with transport coefficients (for heat and momentum) that are consistent with our present understanding of strongly stratified shear-induced turbulence \citep{Chinietal22,Shahal24}. It is important to acknowledge, however, that the model is still incomplete, and that many of the original problems suffered by the seminal discussion of SZ92 still remain.  The purpose of our investigation was to shed light on the dynamical issues raised, and to offer a self-consistent model as a consequence, not necessarily to advocate that that model fundamentally explains the Sun.  
For completeness we now summarize some of the remaining issues that need to be resolved.

First, we emphasize that the new solution  continues to be contingent on the assumption that the turbulent transport of heat and momentum can be modeled as an anisotropic (diagonal) turbulent diffusion process. However, strongly stratified  turbulence in the tachocline may not {\it necessarily} have diffusive properties \citep{Rhines1975,VallisMaltrud1993,GoughMcIntyre98,Tobiasal2007}. Nor, granted that the tachocline is rotating, would the turbulent viscosity tensor likely to be diagonal \citep[cf.][]{Gough2012}. The extent to which these matters influence the dynamics  remains an open question. 

Assuming the turbulence does indeed act diffusively, a second question concerns the applicability of the stratified turbulence model of \citet{Chinietal22} and \citet{Shahal24} to the solar tachocline. Indeed, these papers focused on modeling non-rotating, strongly stratified, shear-driven hydrodynamic  turbulence. However, the Rossby number $\Delta \Omega / \Omega_\odot$ in the tachocline is small, suggesting that rotation could have a significant impact both on the shear instability itself \citep[cf.][]{Watson81,Garaud01}, and on the turbulence resulting from the shear instability \citep[see, e.g.][]{WAITE_BARTELLO_2006,Pouquetal2018,VanKanAlexakis2022}. 
Furthermore, it is likely that the tachocline is at least somewhat magnetized by the solar dynamo, which would also modify the properties of the turbulence \citep{Tobiasal2007}, and perhaps stabilize, or even further destabilize, the shear instability. 
It will therefore be important in the near future to extend the turbulence model (and its numerical validation) to rotating\, magnetized, stratified shear flows.  

Other criticisms of the SZ92 model also continue to apply here. Notably, while this type of turbulent hydrodynamic model can answer the question of how to quench the latitudinal shear communicated by the convection zone across a thin layer, it cannot explain why the entire radiative zone seems to be in a state of almost-uniform rotation (or why the rigid rotation rate is not significantly more rapid) despite having been spun down, initially quite rapidly, over the past 4.6Gyr. 
Indeed, the solar-wind torque exerted on the surface of the  convection zone is readily communicated to the tachocline by the turbulent convection itself, yet  
the deep interior of the Sun would continue to rotate rapidly 
in the absence of additional mechanisms to transport angular momentum radially outward. 

A magnetic field appears to provide the  most likely explanation, promoting  rapid angular momentum transport along radial field lines in the deep interior without causing at the same time significant compositional transport (which would be incompatible with helioseismic inversions). Several possible magnetic models attempting to explain the uniform rotation of the solar radiative zone have been proposed over the last 25 years, which can loosely be categorized into two classes. In one class of models, the inevitable primordial magnetic field  
\citep{MestelWeiss87,Gough90,RudigerKitchatinov97,MacGregorCharbonneau99}  must be confined almost entirely to the radiative zone below the tachocline
\citep{GoughMcIntyre98,GaraudGaraud08,Strugarek-etal11,AAal13}, penetrating the tachocline in only a latitudinally narrow range in which the vertical shear vanishes, for otherwise field lines directly connected with the base of the convection zone would promote (rather than hinder) the propagation of the latitudinal shear into the interior. In the second class of models, the magnetic field is generated by a dynamo located in the convection zone (or near the top of the tachocline), and diffuses into the radiative interior from above \citep{FDP2001,Barnabeal17,Matilsky-etal22}. Strong turbulent magnetic diffusion in the  tachocline, or  some degree of aperiodicity \citep{Garaud99} or asymmetry \citep{Matilsky-etal24} in the dynamo field structure are required to allow the field to penetrate beyond a shallow skin depth and influence the radiative zone at depth.  

Given the aforementioned uncertainties on the nature of turbulence in the tachocline, as well as the likely role of 
magnetic stresses on the tachocline and radiative interior dynamics, it is clear that the simplistic model discussed in this paper is incomplete, and perhaps hardly relevant to the Sun. However,
we believe that it still contributes both to our understanding of  tachocline confinement and more generally to the field of stellar astrophysics, in several ways. First of all, given the historical significance of the SZ92 model for solar interior physics, we felt that it was important to scrutinize it closely, and after identifying their inconsistent approximation of the thermal energy equation, to correct that problem. After all, it is only by building on solid foundations that more complex models can later be created.  Second, our analysis formally demonstrates that the horizontal transport of heat by turbulent eddies in stellar radiative interiors really ought to be taken into account if the turbulence is already invoked to transport angular momentum horizontally on a fast timescale.  
 This is an important conclusion, because the rotational mixing model proposed by   \citet{Zahn92} that is used in many stellar evolution codes \citep[such as MESA, see][]{Paxton13} assumes fast turbulent horizontal momentum transport but  
ignores the corresponding horizontal heat transport. The mixing model's successor, by \citet{MathisZahn04}, does take it into account and in that respect should perhaps be preferred. 
Finally, it will be interesting to see if a self-consistent model of the whole  radiative interior of the Sun can be created by combining the turbulent hydrodynamical model of the tachocline described here with the magnetized model of the deeper radiative interior advocated by \citet{GoughMcIntyre98}. This is one of the future research avenues we plan to explore.

\begin{acknowledgements}
 This work was funded by NSF AST 2408025, NSF AST 2202253, and NASA grant 80NSSC22M0162. The authors thank Joseph Pedlosky and Geoffrey Vallis for invaluable discussions at the WHOI GFD summer program 2024, as well as Nicholas Brummell and Juri Toomre. 
\end{acknowledgements}

\section*{Appendix: Boundary layer scalings}

In this appendix, we use boundary layer analysis to demonstrate that the time derivative in  (\ref{eq:temp_cart}) cannot be ignored in the regime discussed in Section \ref{sec:newmodel}. We also prove that the tachocline is indeed in thermal-wind  balance, as assumed. As in the main text, we begin by assuming that the dynamics is governed by equations (\ref{eq:mom1})-(\ref{eq:goveqs_cart}), together with (\ref{eq:fluxes_cart}) and (\ref{eq:rscart}).
We also assume for simplicity,  as in Section \ref{sec:constnucart}, that the turbulent diffusivities $\tilde{\nu}_h$, $\tilde{\nu}_v$, $\tilde{\kappa}_h$ and $\tilde{\kappa}_v$ are constant.  

We non-dimensionalize these equations  using $r_{cz}$ as the unit length, $\Delta \Omega^{-1}$ (the horizontal shearing rate) as the unit time, so $r_{cz} \Delta \Omega$ is the unit velocity. We also use $r_{cz} T_m N_m^2/g$ as the unit temperature and $\rho_m r_{cz}^2 \Delta \Omega^2$ as the unit pressure. In these units, the governing equations become 
\begin{eqnarray}
\frac{\partial^2 \psi}{\partial t \partial y} = - \frac{\partial p}{\partial z} + \frac{T}{Fr^2} +  \frac{1}{Re_v} \frac{\partial^3 \psi}{\partial z^2 \partial y} + \frac{1}{Re_h}   \frac{\partial^3 \psi}{\partial y^3 } , \\
- \frac{\partial^2 \psi}{\partial t \partial z} + \frac{u}{Ro} = -\frac{\partial p}{\partial y} - \frac{1}{Re_v} \frac{\partial^3 \psi}{\partial z^3} - \frac{1}{Re_h}   \frac{\partial^3 \psi}{\partial y^2\partial z } ,
\\
 \frac{\partial u}{\partial t} + \frac{ 1}{Ro} \frac{\partial \psi}{\partial z} = \frac{1}{Re_v} \frac{\partial^2 u}{\partial z^2} + \frac{1}{Re_h}   \frac{\partial^2 u}{\partial y^2 } ,  \label{eq:azimuth_cart_nondim} \\ 
  \frac{\partial T}{\partial t} +  \frac{\partial \psi}{\partial y}  =    \frac{1}{Pe_v} \frac{\partial^2 T}{\partial z^2} + \frac{1}{Pe_h}  \frac{\partial^2 T}{\partial y^2 },  \label{eq:temp_cart_nondim}
\end{eqnarray}
where we have defined the following non-dimensional numbers: 
\begin{eqnarray}
Re_h = \frac{r_{cz}^2 \Delta \Omega}{\tilde\nu_h}, Re_v = \frac{r_{cz}^2 \Delta \Omega}{\tilde\nu_v},  \\ 
Pe_h = \frac{r_{cz}^2 \Delta \Omega}{\tilde \kappa_h}, Pe_v = \frac{r_{cz}^2 \Delta \Omega}{\tilde \kappa_v}, \\  Ro = \frac{\Delta \Omega}{2\Omega_\odot}, 
Fr = \frac{\Delta \Omega}{N_m}.
\end{eqnarray}
In what follows, we focus on the case where
\begin{equation}
    Fr \ll Ro \ll 1,
\end{equation}
which is consistent with the fact that $\Delta \Omega \ll \Omega_\odot \ll N_m$ in the solar tachocline. 
We also assume that both $Re_h$ and $Pe_h$ are $O(1)$, to capture the idea that the horizontal length scale of turbulent eddies is $O(r_{cz})$. By contrast we have $Pe_v, Re_v \gg 1$, because $\tilde \nu_v \ll \tilde \nu_h$, and $\tilde \kappa_v \ll \tilde \kappa_h$ in strongly stratified turbulence.

Anticipating that the tachocline is thin, we now rescale the vertical coordinate as  $\zeta = z/ \alpha$, where $\alpha \ll 1$ is a small number to be determined from the asymptotic analysis. We then obtain
\begin{eqnarray}
 \frac{\partial^2 \psi}{\partial t \partial y} = - \frac{1}{\alpha} \frac{\partial p}{\partial \zeta} + \frac{T}{Fr^2} +  \frac{1}{\alpha^2 Re_v} \frac{\partial^3 \psi}{\partial \zeta^2 \partial y} + \frac{1}{Re_h}   \frac{\partial^3 \psi}{\partial y^3 },  \\
 \frac{\partial^2 \psi}{\partial t \partial \zeta} - \frac{\alpha u}{Ro} = \alpha \frac{\partial p}{\partial y} + \frac{1}{\alpha^2 Re_v} \frac{\partial^3 \psi}{\partial \zeta^3} + \frac{1}{Re_h}   \frac{\partial^3 \psi}{\partial y^2\partial \zeta } ,
\\
 \frac{\partial u}{\partial t} +  \frac{ 1}{\alpha Ro}\frac{\partial \psi}{\partial \zeta}  = \frac{1}{\alpha^2 Re_v} \frac{\partial^2 u}{\partial \zeta^2} + \frac{1}{Re_h}   \frac{\partial^2 u}{\partial y^2 } ,   \\ 
  \frac{\partial T}{\partial t} +  \frac{\partial \psi}{\partial y}  =    \frac{1}{\alpha^2 Pe_v} \frac{\partial^2 T}{\partial \zeta^2} + \frac{1}{Pe_h}  \frac{\partial^2 T}{\partial y^2 }  .
\end{eqnarray}
In Section \ref{sec:newmodel} we have assumed that the vertical diffusion of momentum and temperature are both negligible compared with the horizontal diffusion. Mathematically, we see that this assumption 
requires  
$\alpha^2 Re_v \gg 1$ and $\alpha^2 Pe_v \gg 1$. Self-consistency of this assumption must be verified a posteriori (see below). If it holds, then the dominant balance in each equation is: 
\begin{eqnarray}
   \frac{\partial^2 \psi}{\partial t \partial y} = - \frac{1}{\alpha} \frac{\partial p}{\partial \zeta} + \frac{T}{Fr^2} +   \frac{1}{Re_h}   \frac{\partial^3 \psi}{\partial y^3 }, \label{eq:hydrostatics}\\
 \frac{\partial^2 \psi}{\partial t \partial \zeta} - \frac{\alpha u}{Ro} = \alpha \frac{\partial p}{\partial y} + \frac{1}{Re_h}   \frac{\partial^3 \psi}{\partial y^2\partial \zeta } \label{eq:heliostrophic} ,
\\  
 \frac{\partial u}{\partial t} +  \frac{ 1}{\alpha Ro}\frac{\partial \psi}{\partial \zeta}  = \frac{1}{Re_h}   \frac{\partial^2 u}{\partial y^2 } ,  \label{eq:azimuth_cart_nondim}  \\ 
  \frac{\partial T}{\partial t} +  \frac{\partial \psi}{\partial y}  = \frac{1}{Pe_h}  \frac{\partial^2 T}{\partial y^2 } \label{eq:temp_cart_nondim} .
\end{eqnarray}
Inspection of equation (\ref{eq:temp_cart_nondim}) shows that in this regime the horizontal temperature diffusion term and the time derivative term are both $O(T)$ (recalling that $Pe_h = O(1)$). The time-derivative is therefore important and cannot be ignored. This equation also reveals that $T = O(\psi)$. Similarly, the time derivative term in the azimuthal component of the momentum equation (\ref{eq:azimuth_cart_nondim}) is of the same order as the corresponding horizontal momentum diffusion term, and must therefore be kept. Furthermore, because the unit velocity was selected so that the azimuthal flow $u=O(1)$, we therefore also have $\psi = O(\alpha Ro)$ from the same equation, revealing that $\psi \ll 1$.

Using this information in equation (\ref{eq:heliostrophic}), we see that the Coriolis term is $O(Ro^{-2})$ times larger than the time derivative and the diffusion term, thus showing that the dominant balance in this regime is necessarily geostrophic: 
\begin{equation}
    \frac{u}{Ro} \simeq -\frac{\partial p}{\partial y}, \label{eq:app_geostro}
\end{equation}
and that $p=O(1/Ro)$.

Finally, using the fact that $T = O(\psi)$ and $Fr \ll 1$ in (\ref{eq:hydrostatics}) shows that the time derivative and the diffusion terms are both negligible compared with the buoyancy term $T / Fr^2$, so the flow is in hydrostatic equilibrium with 
\begin{equation}
    \frac{1}{\alpha} \frac{\partial p}{\partial \zeta} \simeq \frac{T}{Fr^2}. 
\end{equation}
From this we deduce that $p = O(\alpha T/Fr^2) = O(\alpha^2Ro/Fr^2)$. Combining these estimates reveals 
the size of $\alpha$ to be
\begin{equation}
    \alpha = O\left(\frac{Fr}{Ro}\right) = O\left(\frac{2\Omega_\odot}{N_m}\right). 
\end{equation}
This means that the tachocline is indeed in thermal-wind balance and has a characteristic vertical length scale $h = O((\Omega_\odot/N_m)r_{cz})$, which recovers our findings from Section \ref{sec:constnucart}.

In order for the tachocline to be thin, and for this boundary layer scaling to be self-consistent, $\alpha$ must satisfy 
\begin{equation}
\alpha \ll 1, \alpha^2 Re_v \gg 1, \mbox{ and } \alpha^2 Pe_v \gg 1.
\label{eq:case1const}
\end{equation}
The first of these conditions is automatically satisfied because $Fr \ll Ro$. The other two conditions are equivalent to (\ref{eq:selfconnew}) and their validity depends on the turbulence model used. In the \citet{Chinietal22} model for instance, 
\begin{equation}
    Re_v \simeq Fr^{-3/2} Re_h \ll Pe_v \simeq Fr^{-2} {Pe_h}.
\end{equation}
We must therefore verify that 
\begin{equation}
    \alpha^2 Re_v \simeq  \frac{Fr^{1/2}}{Ro^2} Re_h \gg 1 \rightarrow Fr \gg Ro^4. 
\end{equation}
It is easy to check that this is equivalent to (\ref{eq:selfconst_main}), which is satisfied in the solar tachocline.

\bibliography{Biblio}

\end{document}